\documentclass[10pt]{article}
\usepackage{graphicx}

\def\Title#1{\begin{center} {\Large #1 } \end{center}}
\def\Author#1{\begin{center}{ \sc #1} \end{center}}
\def\Address#1{\begin{center}{ \it #1} \end{center}}

\newcommand\pubblock{\rightline{\begin{tabular}{l} Proceedings of the Fifth Annual LHCP\\ \pubnumber\\
         \pubdate  \end{tabular}}}

\newenvironment{Abstract}{\begin{quotation} \begin{center} 
             \large ABSTRACT \end{center}\bigskip 
      \begin{large}}{\end{large} \end{quotation}}

\newenvironment{Presented}{\begin{quotation} \begin{center} 
             PRESENTED AT\end{center}\bigskip 
      \begin{center}\begin{large}}{\end{large}\end{center} \end{quotation}}

\def\Acknowledgements{\bigskip  \bigskip \begin{center} \begin{large}
             \bf ACKNOWLEDGEMENTS \end{large}\end{center}}

\def\beq{\begin{equation}}
\def\eeq#1{\label{#1}\end{equation}}
\def\eeqn{\end{equation}}

\def\beqa{\begin{eqnarray}}
\def\eeqa#1{\label{#1}\end{eqnarray}}
\def\eeqan{\end{eqnarray}}

\let\bar=\overbar

\def\Dslash{\not{\hbox{\kern-4pt $D$}}}
\def\dslash{\not{\hbox{\kern-2pt $\del$}}}

\def\msb{{\bar{\ssstyle M \kern -1pt S}}}

\usepackage{color}
\usepackage{subfigure}
\usepackage[export]{adjustbox}
\usepackage[bottom]{footmisc}

\newcommand\pubnumber{ ATL-PHYS-PROC-2017-100 }

\newcommand\pubdate{\today}

\def\affiliation{
Department of Physics and Astronomy, University of Pittsburgh\\
3941 O'Hara Street, Pittsburgh, Pennsylvania 15260, United States of America}

\begin{document}

\large
\begin{titlepage}
\pubblock

\vfill
\Title{  DARK MATTER SEARCHES AT THE LHC }
\vfill
\Author{ T.\ M.\ HONG\\
  {\textup{On behalf of the ATLAS and CMS Collaborations}}
}

\noindent\begin{center}\includegraphics[height=35mm]{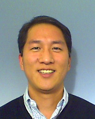}\end{center}
\Address{\affiliation}
\vfill
\begin{Abstract}
{
\noindent
We present a summary of the current status of searches for dark matter at the LHC from the ATLAS and CMS experiments. For various assumptions in the simplified parameter space, the LHC exclusions is complementary to direct detection results. Mono-object analyses in search of dark matter and various analyses searching for dark matter mediators are presented.
}
\end{Abstract}
\vfill

\begin{Presented}
The Fifth Annual Conference\\
 on Large Hadron Collider Physics \\
Shanghai Jiao Tong University, Shanghai, China\\ 
May 15--20, 2017
\end{Presented}
\vfill
\end{titlepage}
\def\thefootnote{\fnsymbol{footnote}}
\setcounter{footnote}{0}
%

\normalsize 

\def\mH{m_H}
\def\mA{m_A}
\def\mZprime{m_{Z^\prime}}
\def\mdm{m_{\mbox{\tiny DM}}}
\def\mdmp{m_{\mbox{\tiny DM}\textrm{-}p}}
\def\gZprime{g_{Z^\prime}}
\def\gq{g_{q}}
\def\gdm{g_{\mbox{\tiny DM}}}
\def\mmed{m_{\rm med}}
\def\mjj{m_{jj}}
\def\met{E^{\rm miss}_{\rm T}}
\def\pT{p_{\rm T}}
\def\ifb{\,\mbox{fb}^{-1}}
\def\cmsq{\,\mbox{cm}^2}
\def\TeV{\,\mbox{TeV}}
\def\GeV{\,\mbox{GeV}}
\def\sigmadm{\sigma_{\mbox{\tiny DM}\textrm{-}p}^{\rm sd}}
\def\sigmaall{\sigma_{\mbox{\tiny DM}\textrm{-}p}}
\def\Binv{\mathcal{B}_{\rm inv}}
\hyphenation{Combined-Summary-Plots}

\section{Introduction}\label{sec:intro}

Searches for dark matter has become one of the most popular topics at the LHC in recent years.
It is widely expected that dark matter interacts with ordinary matter at an energy scale not far from weak interactions; most searches at the LHC loosely rely on this assumption.

During the Run-1 period (2009--2013), LHC searches mainly focused on the effective vertex paradigm as illustrated on the left side of Figure~\ref{fig:zoomin}~\cite{effective}.
The black blob represents the link between dark matter and ordinary matter.\footnote{
  In this proceeding, dark matter candidates are denoted as $\chi$ or DM;
  ordinary matter is denoted as $q$, which represent a quark or fermion, depending on the context;
  DM mediators are denoted as $A$ or ``med,'' unless otherwise noted.
}
The legs that stick out can be labeled to describe annihilations ($\chi\bar\chi\to q\bar q$), scattering ($\chi q\to\chi q$), and production ($q\bar q\to\chi\bar\chi$); see the left side of Figure~\ref{fig:zoomin}.
The last in the list is the focus at the LHC.

More recently, during the Run-2 period (2015--current), simplified models \cite{simplified} became a popular way to resolve the effective vertex as illustrated on the right side of Figure~\ref{fig:zoomin}.
Inserting a propagator into the picture factorizes the $s$-channel diagram with triple-point vertices, $\chi\bar\chi A$ and $q\bar qA$.
New experimental techniques, discussed later, have extended the reach of such searches.

In the case of null results, which is the situation for all searches thus far, assumptions must be made to visualize the exclusions in plots.
In the simplified model the matrix element for the interaction involves four parameters---$\gq$, $\gdm$, $\mdm$, and $\mmed$---as illustrated in the cartoon of Figure~\ref{fig:cartoon}a.
Since the $\sigma_{pp\to\chi\bar\chi}$ is proportional to the number of such events, a null observation can be translated into an exclusion region in a two-dimensional parameter space (say $\mmed$-$\mdm$) can be excluded after assuming two parameters (say $\gq$ and $\gdm$).
For the ideal case, the exclusion is shown in the cartoon of Figure~\ref{fig:cartoon}b as a triangular region below the diagonal line above which the $A\to\chi\chi$ decay is off-shell.

\begin{figure}[b!]
\centering
\includegraphics[width=0.65\textwidth]{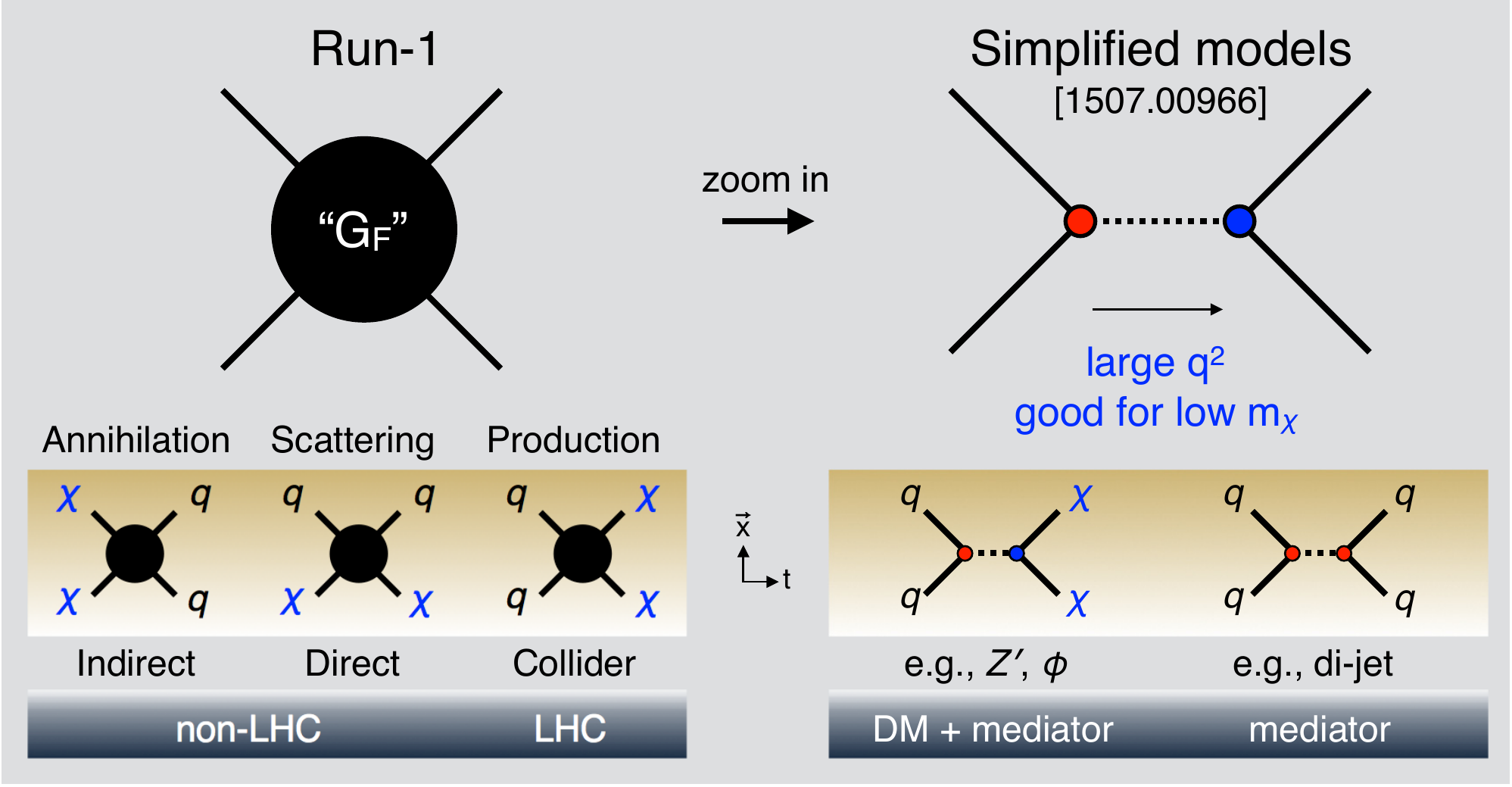}%
\caption{
  Cartoon of the effective vertex framework (left) and the simplified model framework (right).
}
\label{fig:zoomin}
\end{figure}

Furthermore, we ``translate'' the above interpreation into the exclusion of $\sigma_{p\chi\to p\chi}$ in order to compare the LHC results with non-LHC results. 
Such a cross section, e.g., for the spin-dependent interaction, is
\begin{equation}
  \sigmadm \sim \left( \gq\cdot\gdm\cdot \frac{\mdmp}{(\mmed)^2} \right)^2,
\end{equation}
where $\mdmp$ is the reduced mass for the system of proton and dark matter.
Using this relation, the curve in $\mdm$-$\mmed$ space is converted into one in $\mdm$-$\sigmadm$ space by trading $\mmed$ for $\sigmadm$.
Roughly speaking the exclusion is shown in Figure~\ref{fig:cartoon}c as an inverted L shape.
We will see that many of the mono-object results in this framework rules out a region of low $\mdm$; this gives complementary coverage with respect to the direct detection results.
Lastly, the remaining cartoon in Figure~\ref{fig:cartoon}d will be discussed in Section~\ref{sec:mediator}.

This proceeding does not attempt to be comprehensive in any way.
It hopes to give the reader a glimpse of the topic with the few results that were presented in the author's talk at LHCP 2017.

\begin{figure}[t!]
\centering
\subfigure[Simplified model $s$-channel    ]{\includegraphics[width=0.25\textwidth]{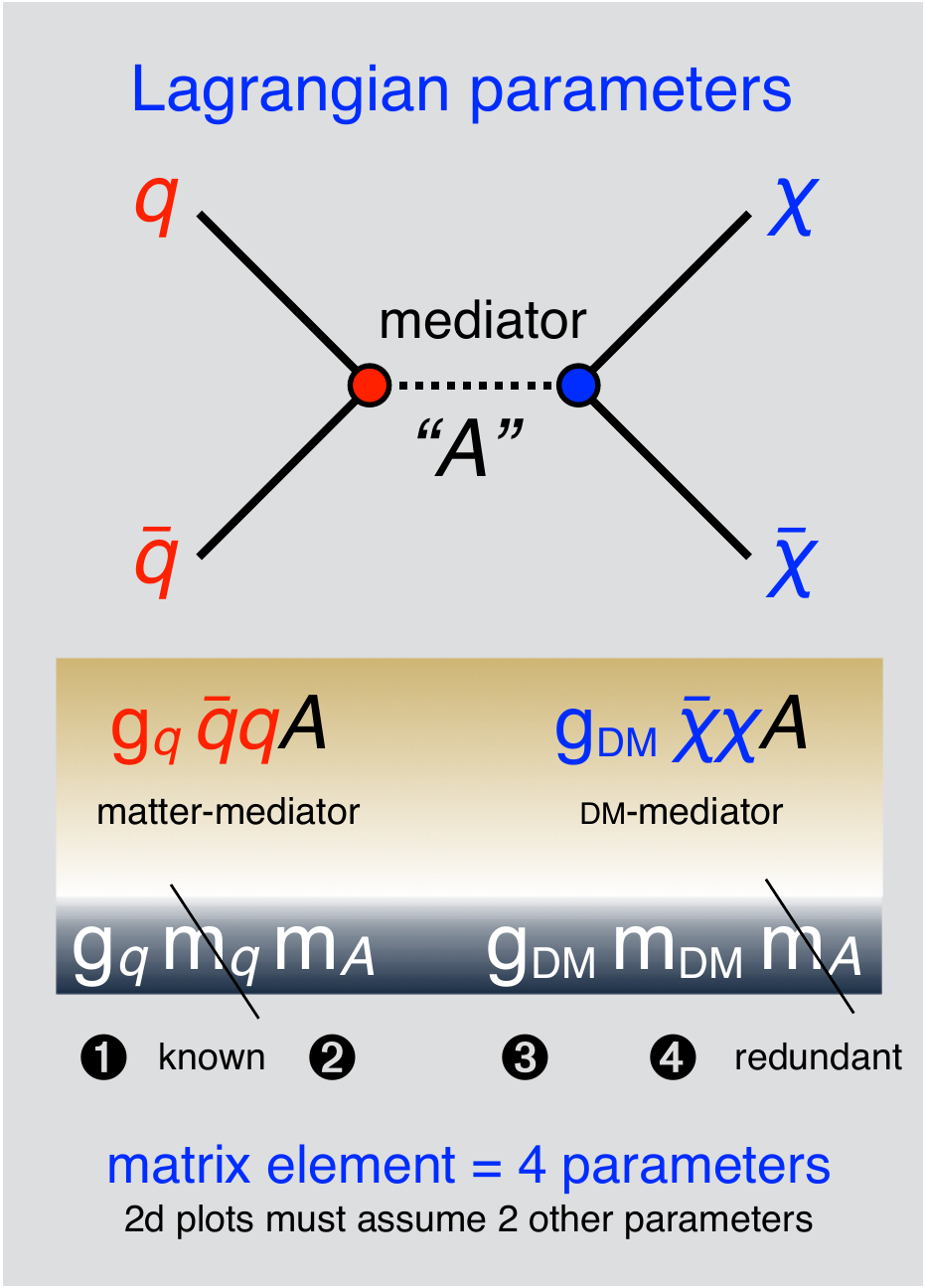}}%
\subfigure[$\mdm$-$\mmed$\,for\,mono-object]{\includegraphics[width=0.25\textwidth]{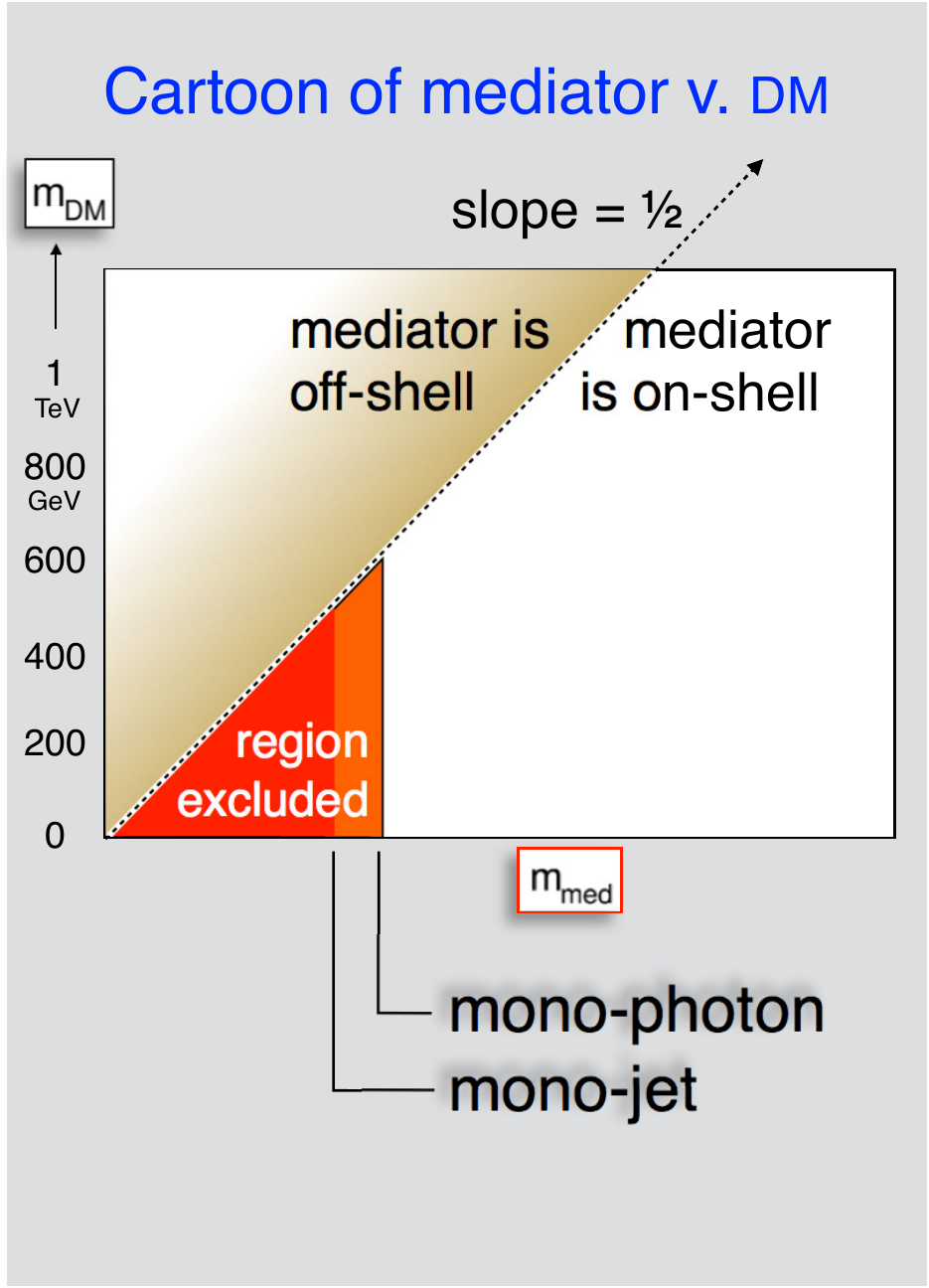}}%
\subfigure[$\mdm$-$\sigmaall$ exclusion    ]{\includegraphics[width=0.25\textwidth]{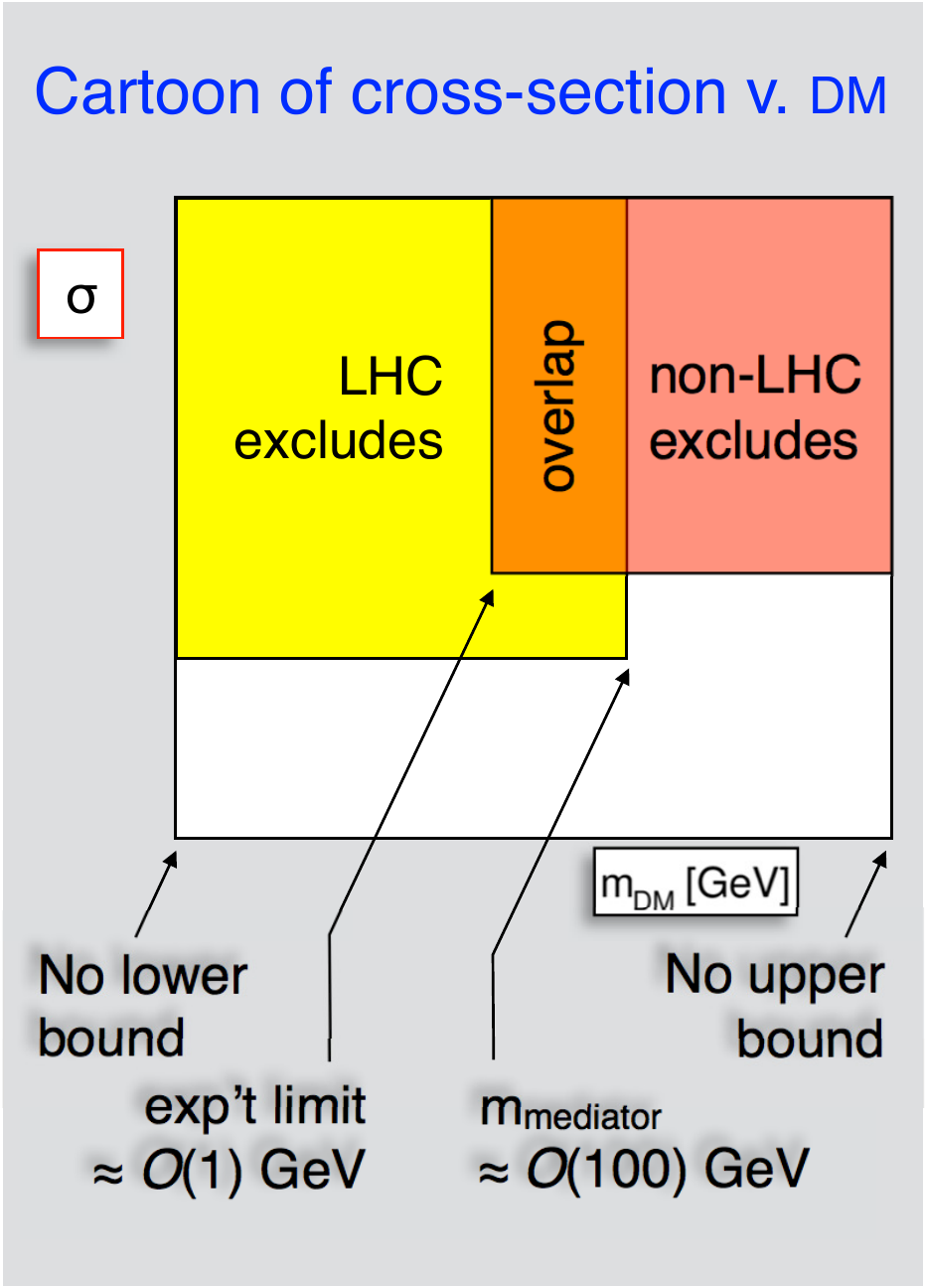}}%
\subfigure[$\mdm$-$\mmed$ for di-jet       ]{\includegraphics[width=0.25\textwidth]{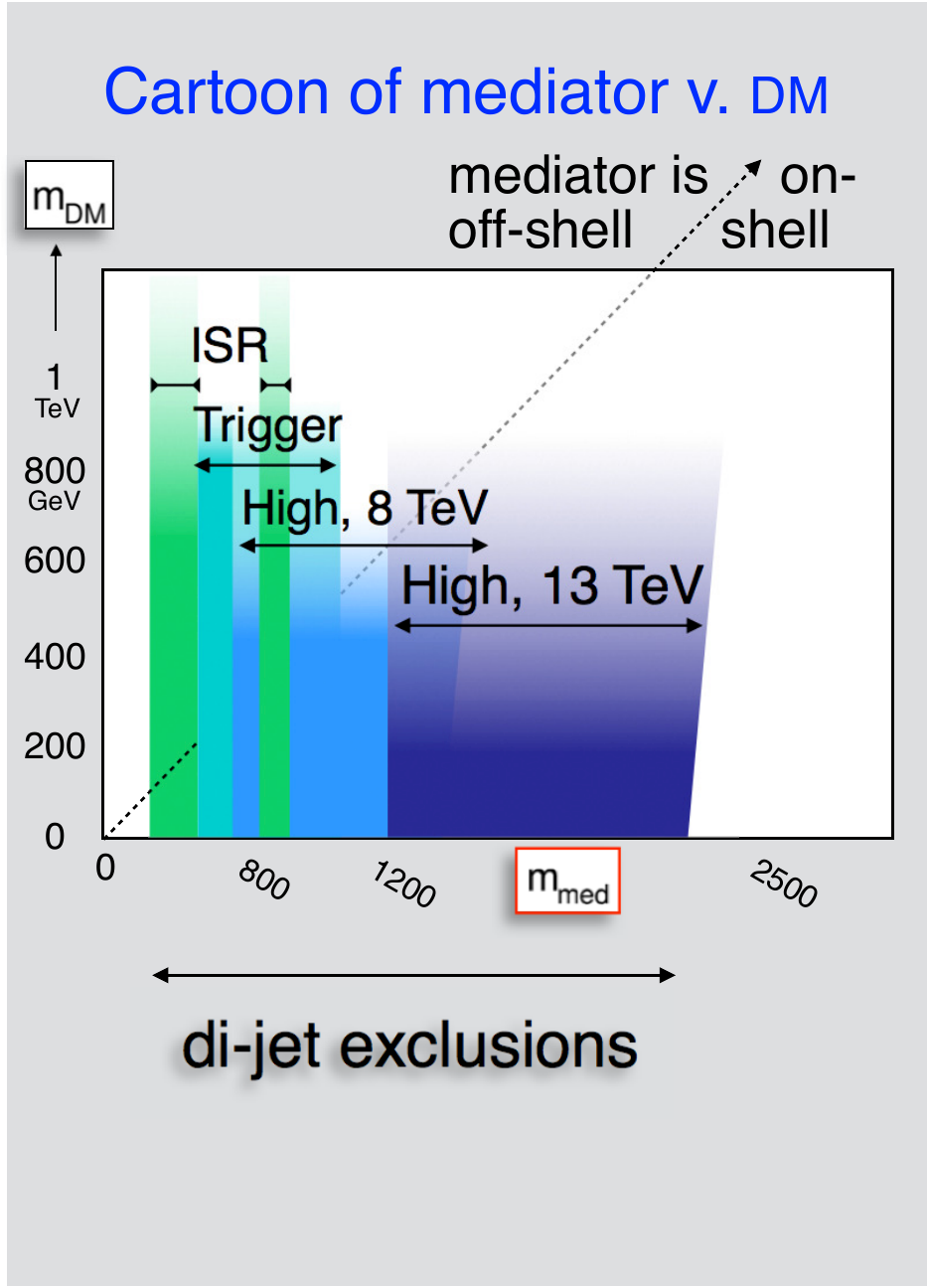}}%
\caption{
  Cartoons to help explain the parametrization (a) and the results (b, c, d).
}
\label{fig:cartoon}
\end{figure}

\section{Mono-object searches for dark matter}\label{sec:dm}

Mono-object\footnote{It should be noted that the nomenclature of ``mono-'' is historical because early searches were focused on mono-jet, where one jet is recoiling against the DM pair. Nowadays it is broadly construed as any {\it system} of observable particles recoiling against the DM pair, so it is a bit of an anachronistic misnomer that we are stuck with.}
analyses are considered the most model-independent searches for dark matter at the LHC.
The targeted interaction is $pp\to\chi\bar\chi+X$, where the $X$ represents the ``mono'' system of observable particles recoiling against the DM pair, $\chi\bar\chi$.
A list for $X$ is, generally speaking, a system composed of jets, photons, weak bosons, Higgs bosons, or heavy flavor quarks ($b$ and $t$), although this list has been growing recently with the increasing number of theoretical ideas.
In this section, three analyses are mentioned to capture the spirit of mono-object searches: the canonical mono-jet analysis and analyses of mono-photon and mono-Higgs.

The mono-jet search looks for a jet recoiling against a DM pair, the latter which manifest itself as $\met$.
Figure~\ref{fig:mono}a shows the $\met$ distribution, where the stacked histogram of background processes is overlaid on various signal models depicted as thick lines.
In the plot, it is notable that the $\met$ distribution for the signal model for the axial-vector mediator of $2\TeV$ mass is flatter than that of the backgrounds, so the signal-to-background ratio ($S$-to-$B$) increases with $\met$.
At lower values of around $200\GeV$ the ratio is $\mathcal{O}(10^{-3})$ and reaches $\mathcal{O}(10^{-1})$ around $800\GeV$.
The small ratios are due to the relatively high cross section values of the production of single weak bosons in association with a hard jet.
For this reason, the $\met$ offline selection in these analyses, which are around $200\GeV$~\cite{monojet}, are generally higher than the lowest online trigger requirement at around $150\GeV$~\cite{trigger}.
Figure~\ref{fig:mono}b shows the event display of one such event with a $\met$ of around $1\TeV$; the jet with $\pT$ of $1\TeV$ is not balanced by anything opposite it in the $r$-$\phi$ cutaway.

The mono-photon search, with a lower expected cross-section with respect to the mono-jet, follows a similar analysis strategy as described above.
As discussed in Section~\ref{sec:intro} and Figure~\ref{fig:cartoon}a, two of the parameters must be fixed in order to exclude a region defined by a curve in a two-dimensional plane.
Figure~\ref{fig:mono}c interprets the null result by excluding a region in the $\mmed$-$\mdm$ plane at 90\% confidence level assuming the coupling values of $\gq=0.25$ and $\gdm=1$ in the Dirac DM model with an axial-vector mediator~\cite{monophoton}.
As described in the introduction, the null result can also be interpreted as a region of exclusion in the $\mdm$-$\sigmadm$ plane by trading $\mmed$ for $\sigmadm$ using Equation 1.

\begin{figure}[p!]
  \centering
  \subfigure[Mono-jet $\met$ distribution \cite{monojet}]{\includegraphics[width=0.4\textwidth]{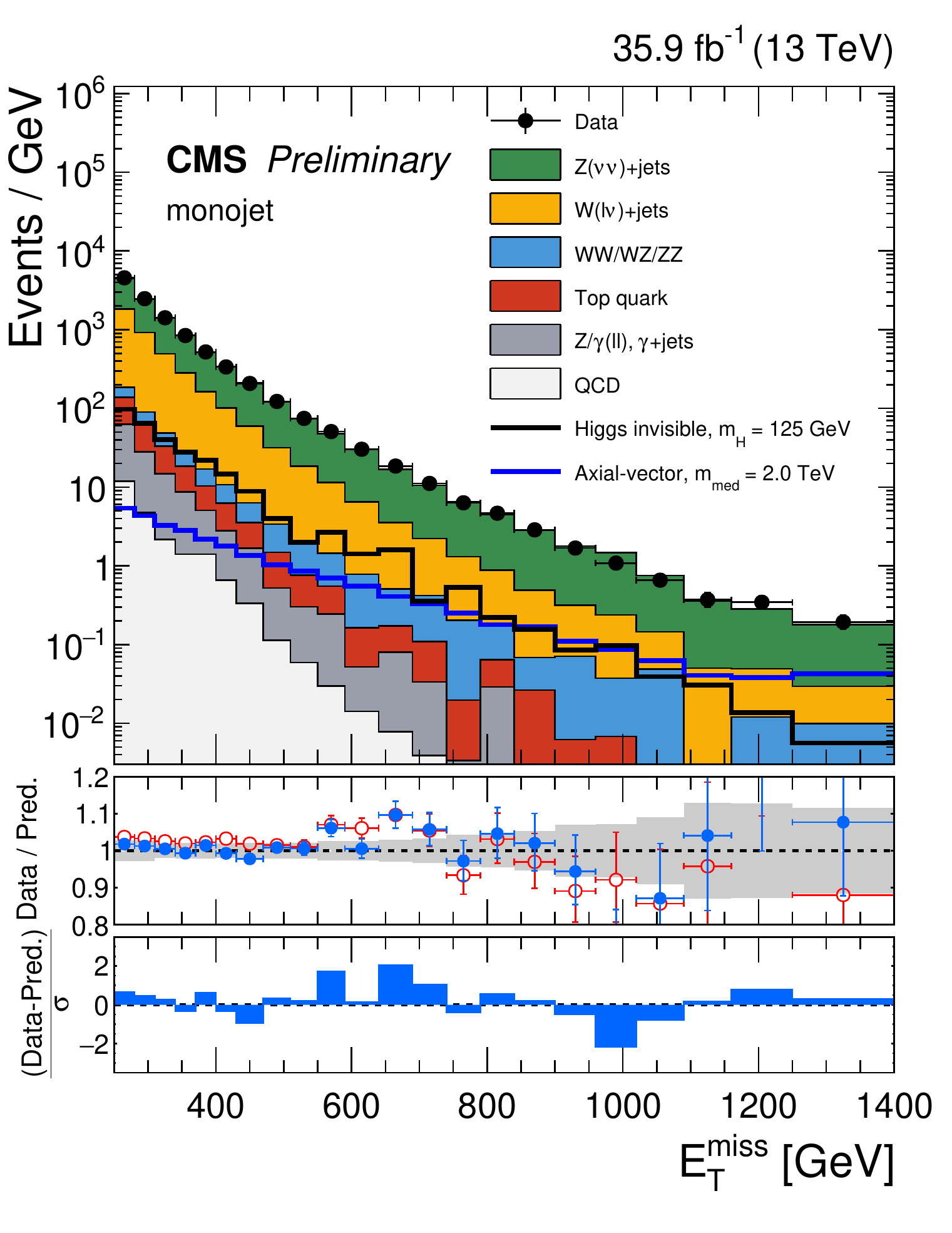}}%
  \hspace{0.35in}\subfigure[
    Mono-jet event display. 
    Jet (downward bars) is balanced by $\met$ (upward arrow), both $1\TeV$
    \cite{eventdisplay}
  ]{\raisebox{0.5in}{\includegraphics[width=0.4\textwidth]{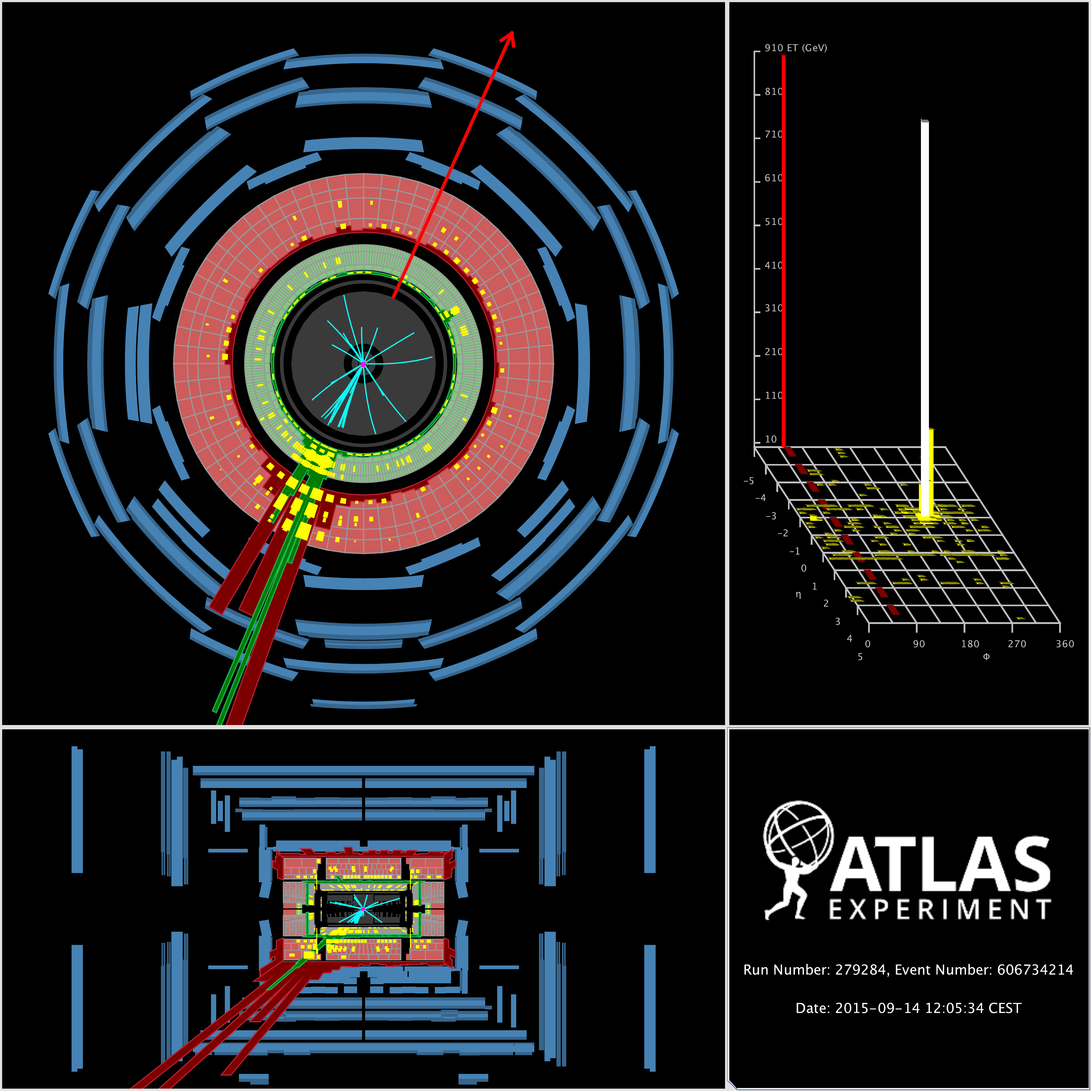}}}\\
  \subfigure[Mono-photon exclusion \cite{monophoton}]{\includegraphics[width=0.500\textwidth]{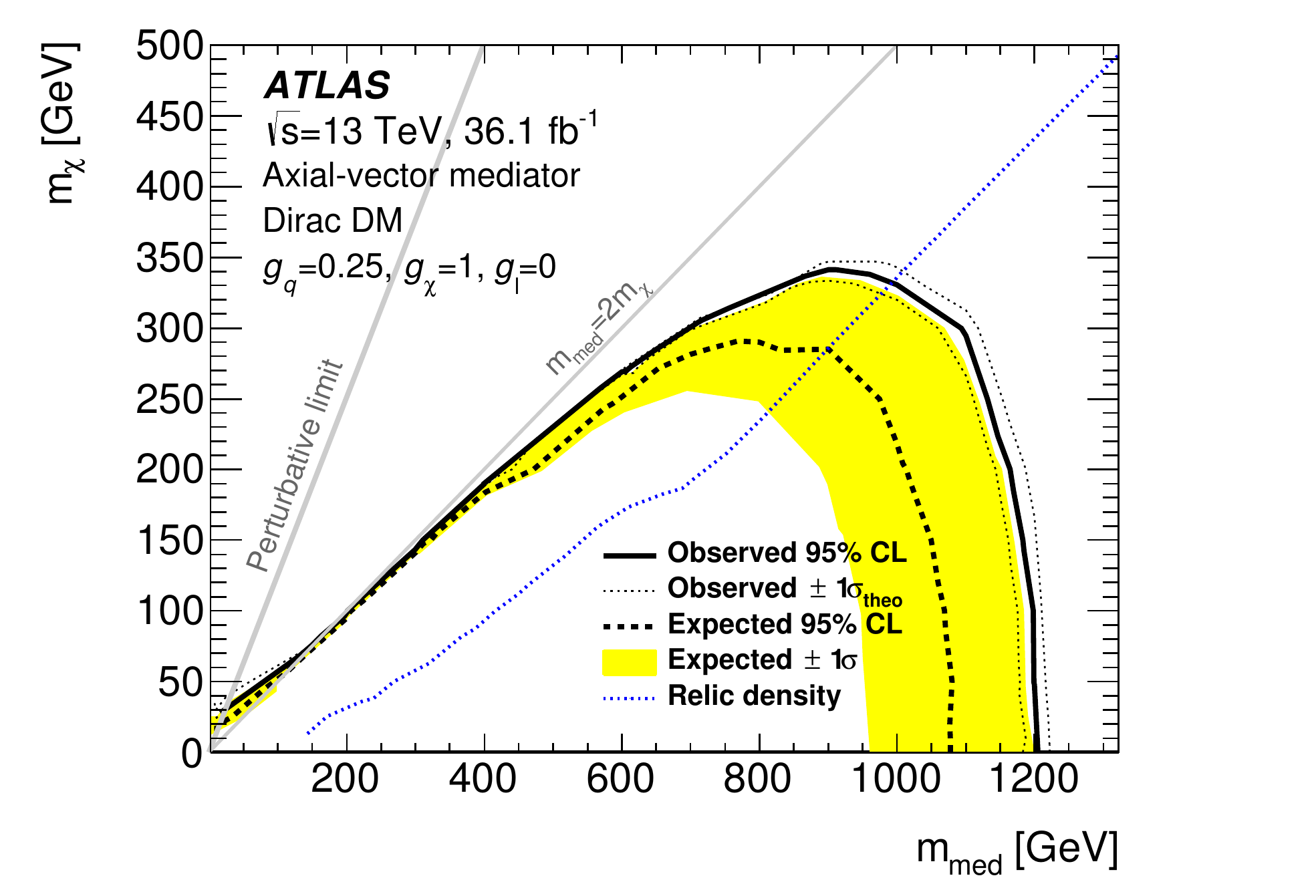}}%
  \hspace{-0.30in}\subfigure[Mono-photon exclusion \cite{monophoton}]{\includegraphics[width=0.500\textwidth]{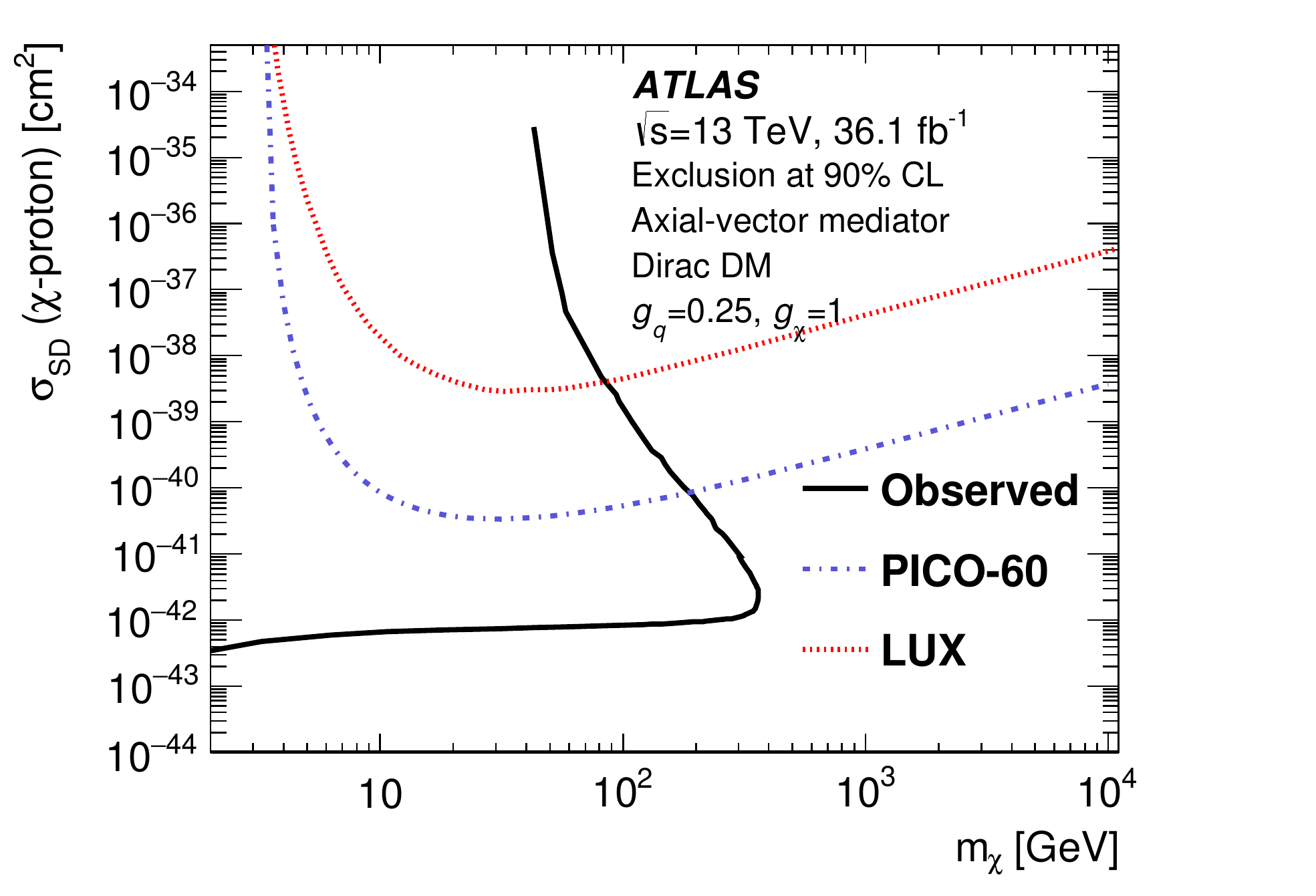}}\\
  \hspace{0.30in}\subfigure[Mono-Higgs diagram]{\raisebox{0.40in}{\includegraphics[width=0.25\textwidth]{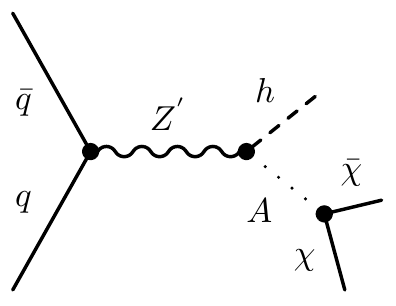}}}%
  \hspace{0.60in}\subfigure[Mono-Higgs exclusion \cite{monoH}]{\includegraphics[width=0.40\textwidth]{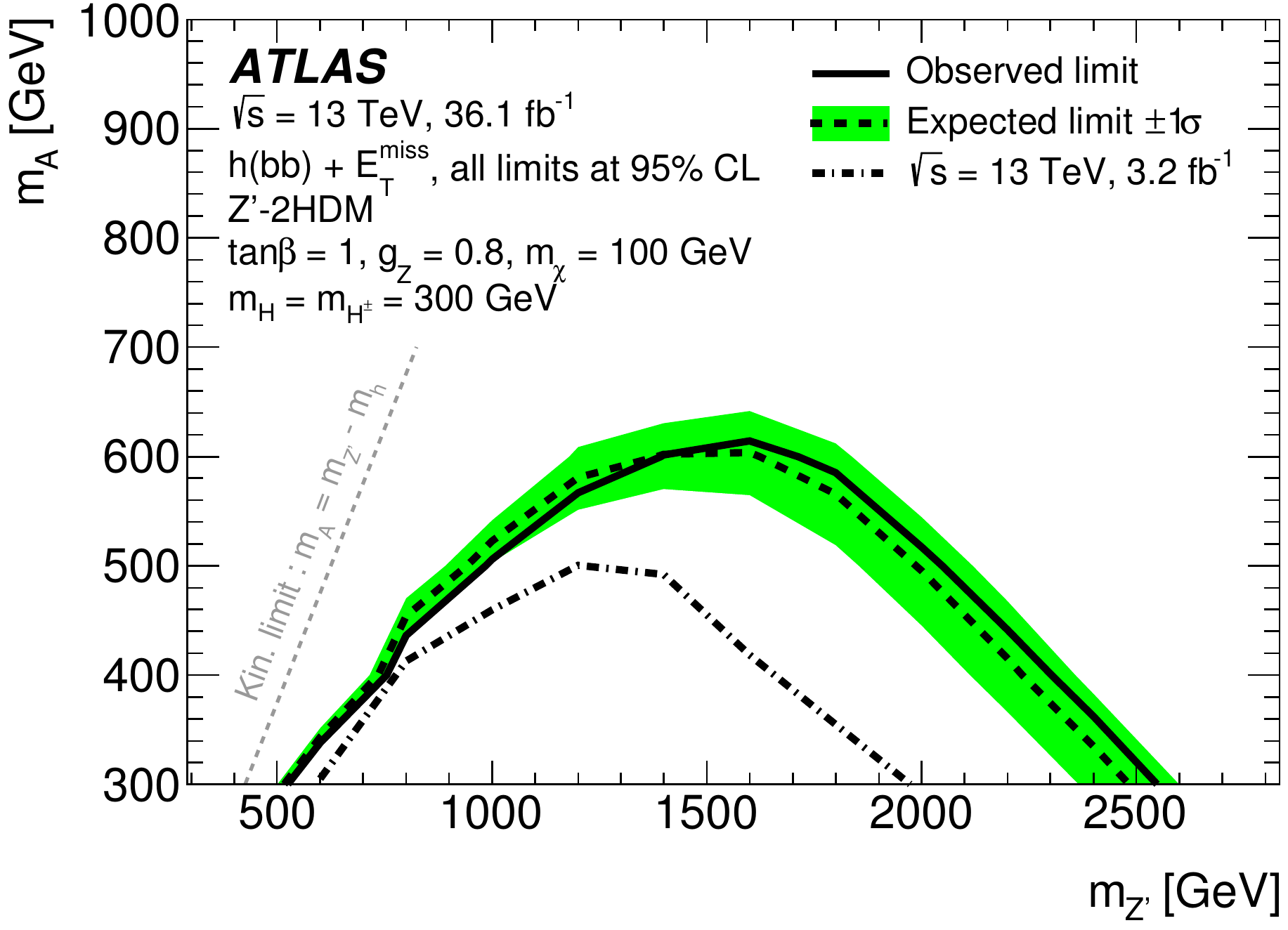}}%
  \caption{
    Mono-object analyses: mono-jet (top row), mono-photon (middle row), mono-Higgs (bottom row).
  }
  \label{fig:mono}
\end{figure}

The mono-Higgs search involves the $s$-channel production of the dark matter mediator $A$ and a Higgs boson; the propagator here is another mediator $Z^\prime$.
Figure~\ref{fig:mono}e shows the Feynman diagram for the process.
Experimentally, the decay channel of $H\to b\bar b$ is chosen for its large branching ratio (around $60\%$) and is fully visible to the detector.
A relatively new technique of ``boosted jets'' is used, wherein the high value of the Higgs boson $\pT$ merges the two $b$-quark jets within a radius (in the $\eta$-$\phi$ plane) of $2\,\mH/\pT$ \cite{monoH}.
The interpretation of the null result requires another layer of complexity because of the additional degrees of freedom in using a Z-prime two Higgs-doublet model ($Z^\prime$-2HDM).
Figure~\ref{fig:mono}f presents the $95\%$ confidence level limit of an excluded region in the plane of $\mA$ vs.\ $\mZprime$, with additional assumptions on $\tan\beta$, $\gZprime$, $\mdm$, and the masses of Higgs-like bosons (all noted in the figure).

This section summarized various aspects of three mono-object analyses in search of dark matter production.
As mentioned in the introduction simplified models have converted the null results into exclusion regions.
With exclusions in the plane of $\mdm$-$\mmed$, the dark matter mediator (or multiple mediators) has become more prominent in the interpration.
These advancements have put into focus not only DM, but on a new sector in which they might live.
The development has lead to searches for mediators, discussed next.

\section{Searches for dark matter mediators}\label{sec:mediator}

The simplified model provides a relatively concise framework in which to expand the interaction between DM and ordinary matter.
In this section, the search for a spin-1 mediator in di-jet and a spin-0 mediator in Higgs decay are discussed.

\begin{table}[b!]
  \caption{
    Features of a DM mediator assuming that it be prompt, colorless, and other simplifications.
    \label{tab:mediator}
  }
  \begin{center}
    \begin{tabular}{l|ll}  
      Features                  & Spin-0 mediator         & Spin-1 mediator \\
      \hline
      Charge                    & $0$ for $s$-channel     & $0$ for $s$-channel \\
      Mass                      & No assumption           & No assumption \\
      Lorentz structure         & Scalar (1), Pseudoscalar ($\gamma_5$) & Vector ($\gamma^\mu$), Axial vector ($\gamma^\mu\gamma_5$) \\
      DM mediator is similar to & Higgs                   & $\gamma$, $Z$, $Z^\prime$ \\
      Coupling ``$g$''          & $\propto$ mass          & $\propto$ charge \\
      Example consequence       & $m_b \gg m_d$           & $Q_b = Q_d$ \\ 
      Example channel           & mono-$b$                & di-jet \\
    \end{tabular}
  \end{center}
\end{table}

In order to get an idea of the characteristics of DM mediators, Table~\ref{tab:mediator} lists some of the features that one might expect.
The list is based on well-established bosons (Higgs, $\gamma$, $Z$) or well-studied hypothetical bosons ($Z^\prime$), so assumptions may not necessarily hold for more complex scenarios.
If the DM mediators does follow the pattern in the table, then, e.g., di-jet is good for spin-1 while mono-$b$ is good for spin-0 because the former couples to the quark's charge whereas the latter couples to the quark's mass.

The di-jet analysis has long been a workhorse in the search for new $Z^\prime$-like resonances and for new contact interactions at a higher energy scale.
It has received renewed attention in light of its possible connection to DM since if the DM mediator couples to ordinary matter (the first small diagram on the right side of Figure \ref{fig:zoomin}), then it can decay back into ordinary matter (the second diagram on the right side).

\begin{figure}[b!]
  \centering
  \subfigure[Distribution of $\mjj$ in dijet searches \cite{dijet}]{\includegraphics[width=0.85\textwidth]{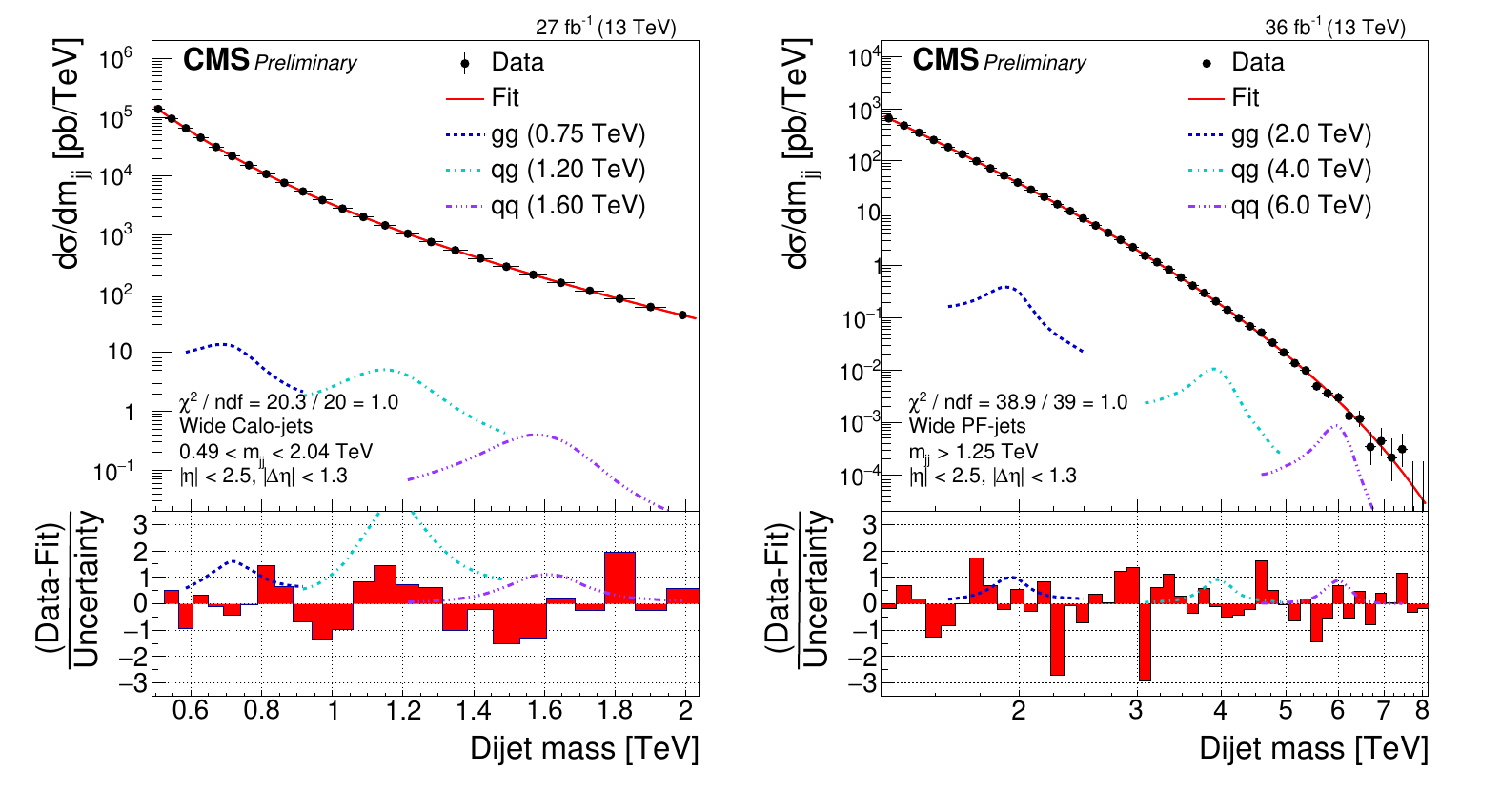}}\\
  \vspace{-9pt}%
  \subfigure[Overlay of axial-vector mediator exclusions from dijet and mono-object searches \cite{summary}]{\includegraphics[width=0.75\textwidth]{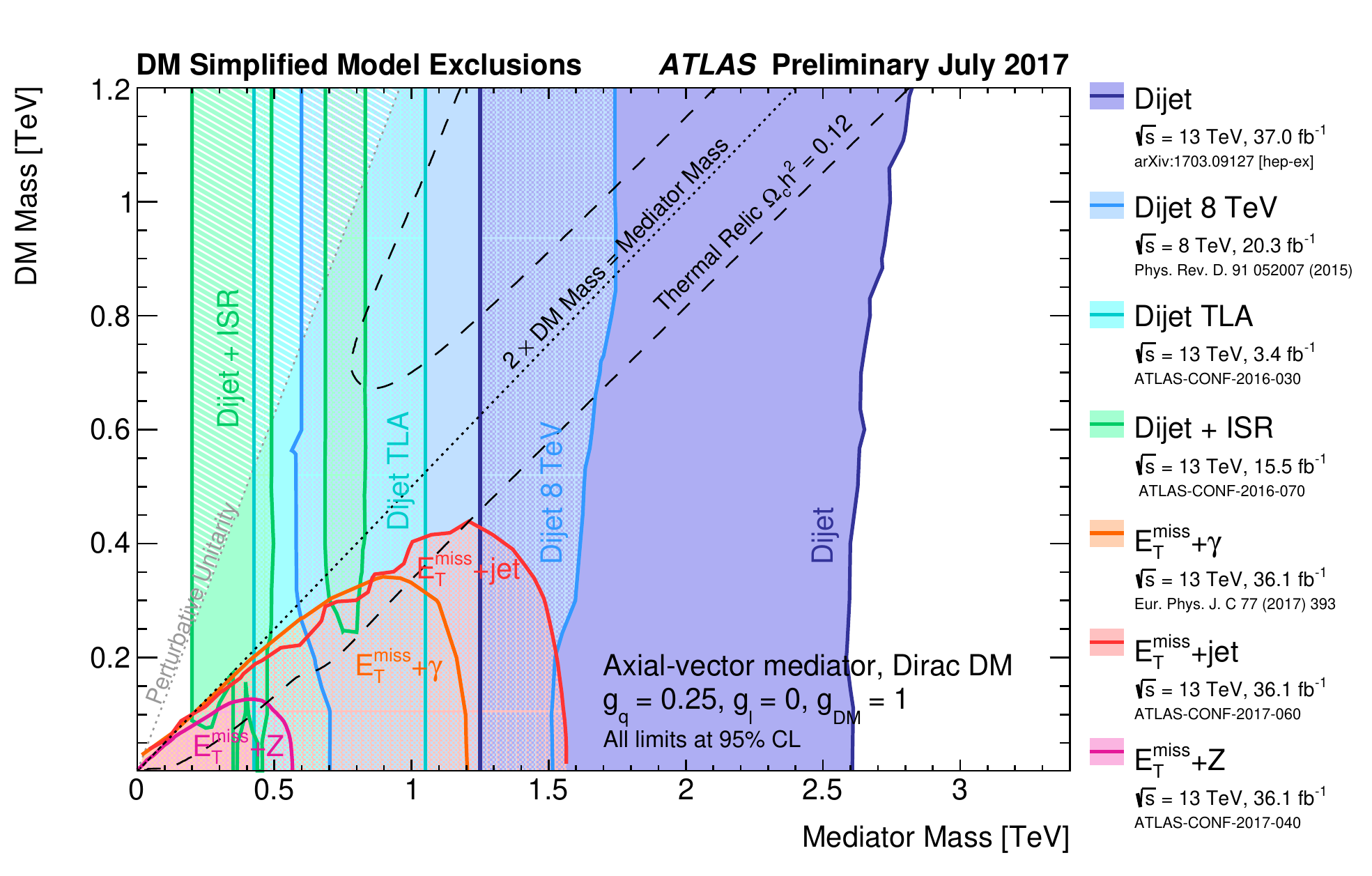}}%
  \caption{
    Di-jet analyses (top) and overlay of axial-vector mediator exclusions (bottom).
    \label{fig:dijet}
  }
\end{figure}

The distribution of the invariant mass of the di-jet system is given in Figures~\ref{fig:dijet}a, spanning lower values from $500\GeV$ to $8\TeV$ \cite{dijet}.
The di-jet analysis has long faced challenges on lowering the threshold, about $2\TeV$ on the di-jet invariant mass because of rate challenges.
The abundant production of jets limits the amount of events with two or more jets one can save to disk.
However, more recently, a new technique of saving partial event information, such as the jet four-vectors and the level of ambient energy, allowed both ATLAS and CMS to save events much below the aforementioned $2\TeV$ threshold; this is shown in Figure~\ref{fig:dijet}a.
There are additional experimental techniques not discussed here---using boosted di-jets and requiring initial-state-radiated (ISR) jet or photon---that have been developed to lower the threshold.

The interpretation of null results of searches for di-jet resonances can be projected in the $\mdm$-$\mmed$ plane with assumptions on $\gq$ and $\gdm$ \cite{summary}.
The cartoon of the situation is given in Figure~\ref{fig:cartoon}d where vertical regions are excluded because the search is sensitive to both on- or off-shell scenarios because the mediator is not decaying to DM but to light quarks with negligible masses.
Figure~\ref{fig:dijet}b overlays the di-jet results and the mono-object results.
The caveat emptor given for the plot is that the exclusion contours are highly dependent on the coupling assumptions.

The last topic in this proceedings is the the Higgs boson.
The Higgs boson has long been a playground for new theoretical ideas \cite{Hexotic};
due to its unique spin-0 status, it has been a stand-in boson for a number of portals to new sectors \cite{portal}, including the dark matter sector.
One idea is to search for an observable $\mathcal{O}(1$--$10\%)$ amount of the branching fraction ($\Binv$) of the Higgs boson decaying to invisible final states.\footnote{%
  The Higgs boson does have a tiny invisible branching fraction at less than $0.1\%$ from $H\to ZZ^\ast \to \nu\bar\nu\nu\bar\nu$.
}

Of the currently established production modes of the Higgs boson, the vector boson fusion (VBF) production gives the best sensitivity on the upper limit on $\Binv$ \cite{Hinv}.
One of the major experimental challenges for this analysis is the trigger,
where either a large value of $\met$, two widely-separated high-$\pT$ jets, or both are required.
On ATLAS, the result relied on the lowest threshold $\met$ requirement at the level-1 calorimeter trigger system, for which the signal is fully efficient at an offline selection of around $150\GeV$.
Because the $\met$ reflects the $\pT$ of the Higgs boson, any increase on the threshold would drastically decrease the signal accordingly.
Figure~\ref{fig:Hinv}a and b shows the $\met$ distribution and $\mjj$ distribution, respectively, for VBF.

Searches in various Higgs production channels are used to put the best combined limit on $\Binv$.
Figure~\ref{fig:Hinv}c shows the CMS $\sigmaall$ exclusion for various models for the observed $90\%$ confidence level upper limit on $\Binv$ at $20\%$ \cite{Hinv2}.
LHC covers region below $\mH/2$, whereas the direct detection covers a region above ${\sim}5\GeV$.
Both are at a similar level of $\sigmaall$ ranging from $10^{-44}$ to $10^{-46}\cmsq$ and have complementary coverage.

\begin{figure}[b!]
  \centering
  \subfigure[Distribution of $\met$ in VBF production \cite{Hinv}]{\includegraphics[width=0.5\textwidth]{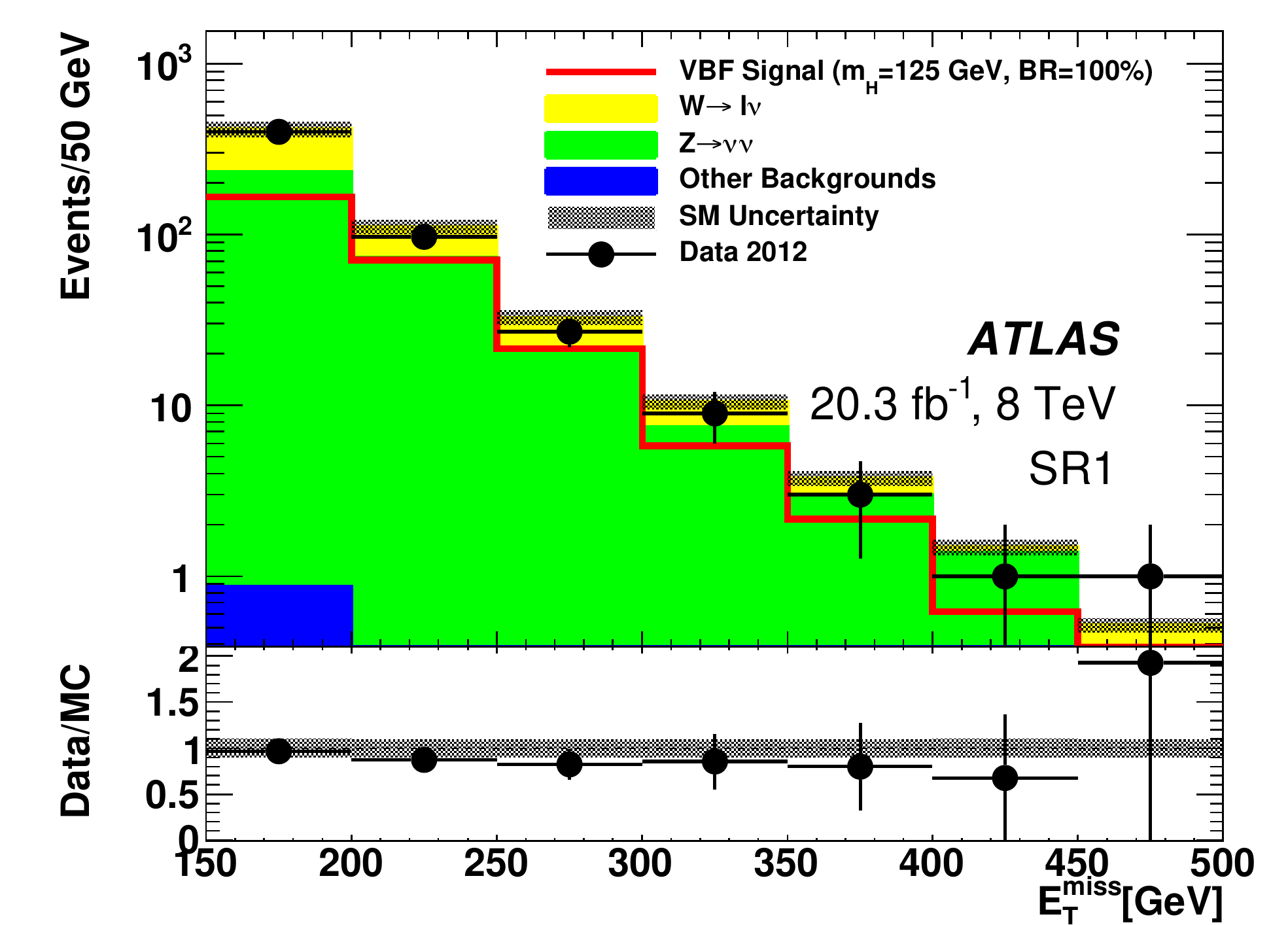}\hspace{0.15in}}%
  \subfigure[Distribution of $\mjj$ in VBF production \cite{Hinv}]{\includegraphics[width=0.5\textwidth]{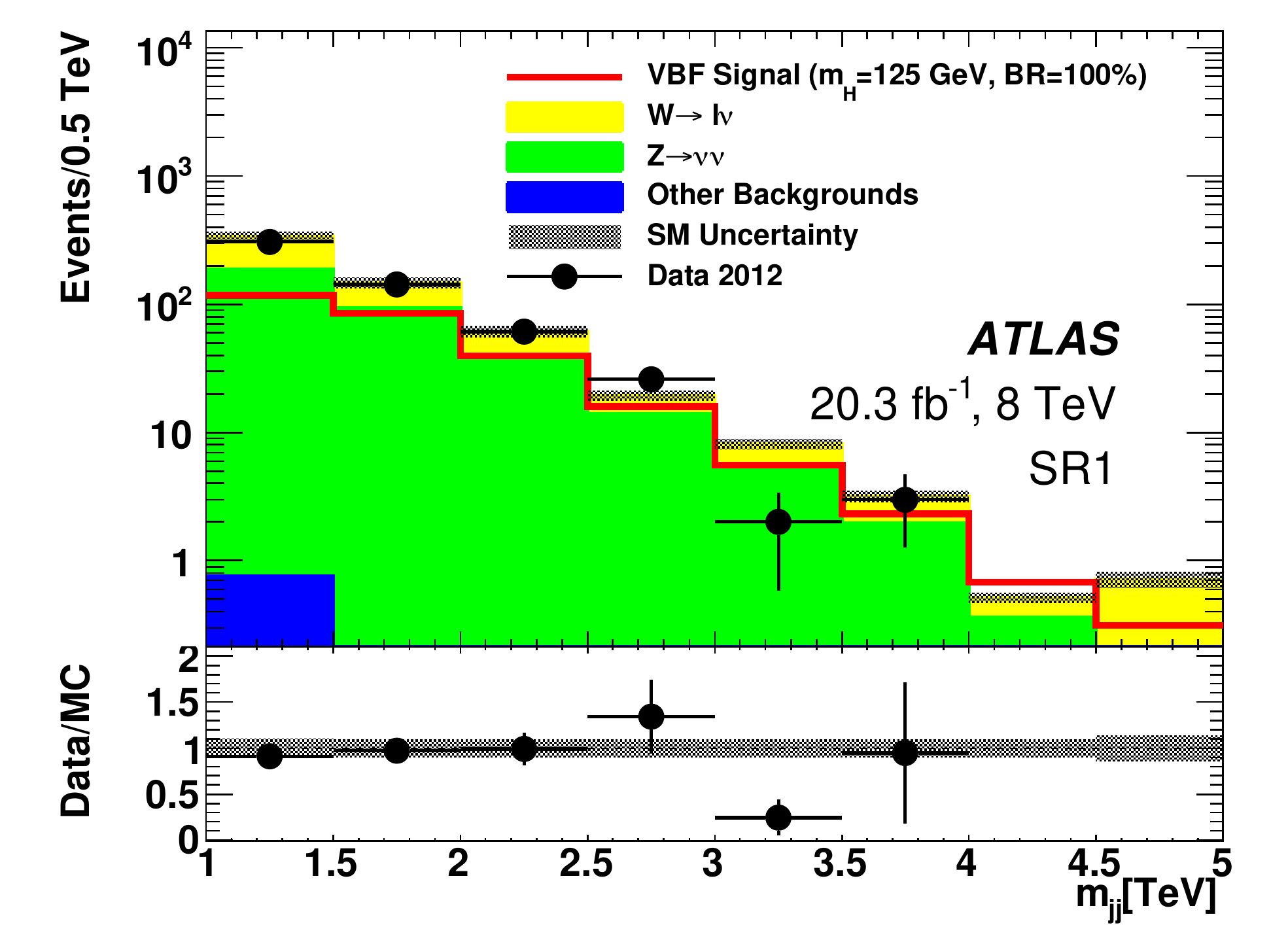}}\\
  \vspace{-9pt}%
  \subfigure[Higgs portal models combining various invisible results \cite{Hinv2}]{\includegraphics[width=0.5\textwidth]{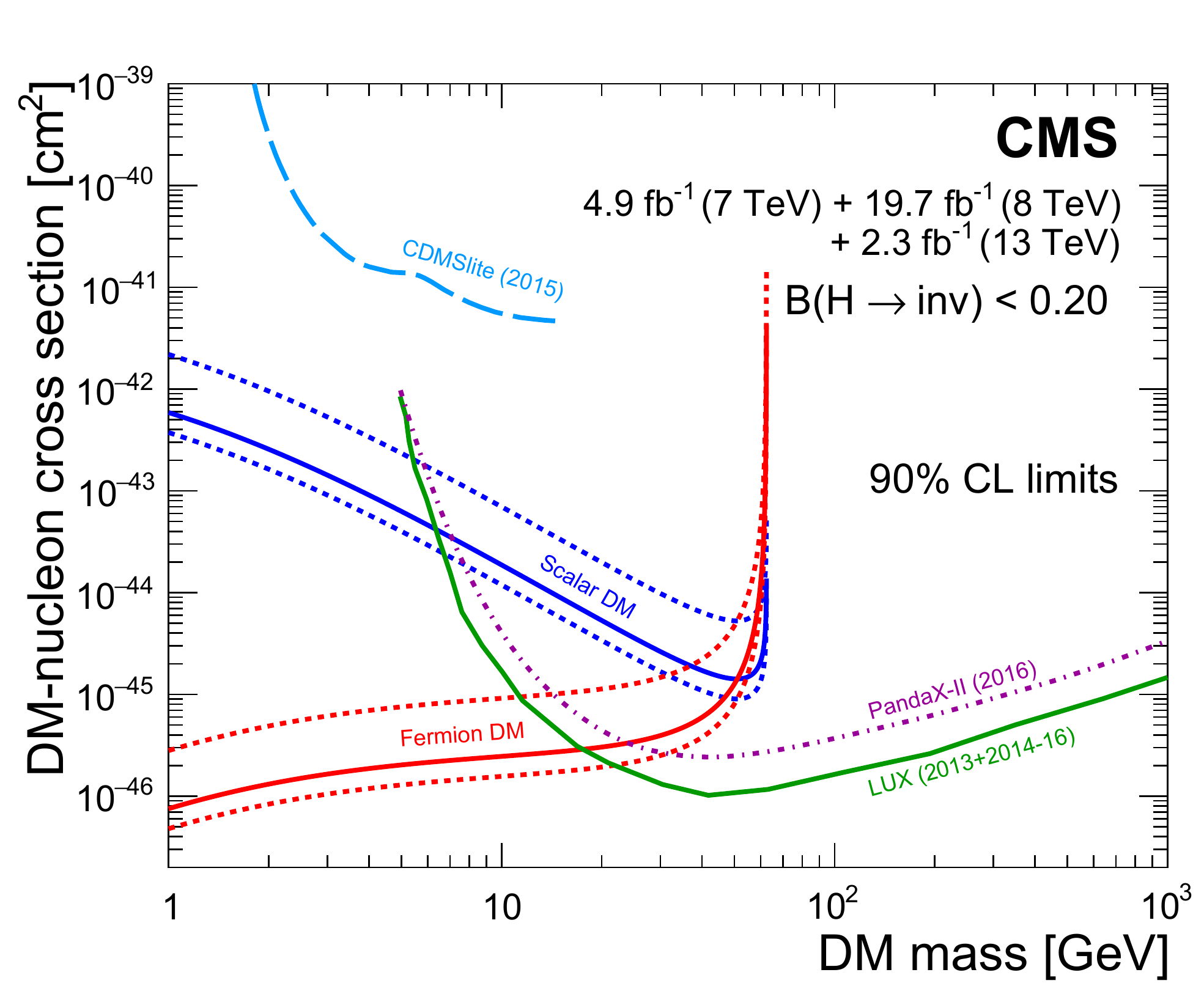}}%
  \caption{
    Higgs to invisible final states: event distributions (top row) and overlay of portal models (bottom).
    \label{fig:Hinv}
  }
\end{figure}

\section{Summary}

A brief overview is given of the LHC searches for dark matter and for mediators that link them to ordinary matter.
The simplified models framework provide a parameter space to interpret the null results.
Among the new experimental handles are ways to deal with the on-going trigger challenges and the use of boosted objects.
We look forward to a rich era of theoretical and experimental productivity in Run-2 and beyond.

\Acknowledgements
I am grateful to B.\ Batell, P.\ Chang, J.\ Duarte, S.\ Meehan, A.\ Tuna, and B.\ Carlson for fruitful discussions.

\end{document}